\documentclass[a4paper,11pt]{article}
\usepackage{amsmath,times,amssymb,latexsym,multicol,graphics}
\usepackage{ascmac,fancybox}
\usepackage{bbm}
\usepackage{wrapfig}
\usepackage{framed}
\usepackage{setspace} 
\usepackage{comment}
\usepackage{authblk}
\usepackage{braket}
\newcommand{\be}{\begin{equation}}
\newcommand{\ee}{\end{equation}}
\newcommand{\nn}{\nonumber}
\newcommand{\D}{\Delta}

\newcommand{\G}{\Gamma}
\newcommand{\la}{\lambda}

\newcommand{\p}{\partial}

\begin{document} 
\begin{titlepage}
\begin{flushright}
{\small OU-HET 910}
 \\
\end{flushright}

\begin{center}

\vspace{1cm}

\hspace{3mm}{\LARGE \bf Geodesic Witten diagrams} \\[3pt] 
\vspace{1mm}
{\LARGE \bf  with an external spinning field}  
 
\vspace{1cm}

\renewcommand\thefootnote{\mbox{$\fnsymbol{footnote}$}}
Mitsuhiro {Nishida}${}^{1}$ and 
Kotaro {Tamaoka}${}^{2}$

\vspace{5mm}

{\small \sl Department of Physics, Osaka University} \\ 
{\small \sl Toyonaka, Osaka 560-0043, JAPAN}

\vspace{5mm}

{\small 
{${}^1$\,nishida@het.phys.sci.osaka-u.ac.jp},  \\
{${}^2$\,k-tamaoka@het.phys.sci.osaka-u.ac.jp}
}

\end{center}

\vspace{5mm}

\noindent
\abstract
We explore AdS/CFT correspondence between geodesic Witten diagrams and conformal blocks (conformal partial waves) with an external  symmetric traceless tensor field. We derive an expression for the conformal partial wave with an external spin-1 field and show that this expression is equivalent to  the amplitude of the geodesic Witten diagram. We also show the equivalence by using conformal Casimir equation in embedding formalism. Furthermore, we extend the construction of the amplitude of the geodesic Witten diagram to an external arbitrary symmetric traceless tensor field. We show our construction agrees with the known result of the conformal partial waves. 

\end{titlepage}
\setcounter{footnote}{0}
\renewcommand\thefootnote{\mbox{\arabic{footnote}}}
\tableofcontents
\flushbottom
\section{Introduction}\label{sec:intro}
Conformal blocks or conformal partial waves (CPW) are fundamental objects in conformal field theory (CFT) and finding their compact expression has long been a research subject of CFT (see, for example, \cite{Ferrara:1972xe, Ferrara:1972ay, Ferrara:1972uq, Ferrara:1973vz, Ferrara:1974nf, Dolan:2000ut, Dolan:2003hv, Dolan:2011dv}). Because of recent progress in the conformal bootstrap program \cite{Rattazzi:2008pe, ElShowk:2012ht}, it has become necessary to obtain better expressions for numerical calculations. In particular, a formula for CPW of external operators with spin such as the stress tensor is important. It will enable us to apply the conformal bootstrap program to various areas, for example, critical phenomena and quantum gravity as the gravity dual of CFT.
 
From the viewpoint of AdS/CFT correspondence \cite{Maldacena:1997re, Gubser:1998bc, Witten:1998qj}, the gravity dual of the conformal partial wave has not been well understood. However, the authors of \cite{Hijano:2015zsa} proposed the correspondence between CPW and geodesic Witten diagrams (GWD) up to normalization based on recent results in the AdS$_3$/CFT$_2$ correspondence of Virasoro conformal blocks \cite{Hartman:2013mia, Asplund:2014coa, Fitzpatrick:2014vua, Fitzpatrick:2015zha, Hijano:2015rla, Alkalaev:2015wia}. GWD are diagrams that represent the scattering process on AdS spacetime, such as the Witten diagram. The difference between GWD and the Witten diagram is the following. In the usual Witten diagram, the interactions are integrated over all points in the bulk. On the other hand, GWD interactions are restricted at geodesics between external operators.  They showed the correspondence between the amplitude of GWD and CPW with four external scalar fields by direct computation and conformal Casimir equation. Moreover, they decomposed the Witten diagrams into GWD. The CPW expansion of the Witten diagrams has been also discussed in \cite{Hoffmann:2000tr,ElShowk:2011ag,Bekaert:2014cea,Bekaert:2015tva}. In \cite{Hijano:2015zsa}, the authors considered the external scalar fields only. The generalization of their results to external fields in arbitrary representation could be useful for the conformal bootstrap program. There have been several developments since \cite{Hijano:2015zsa}, for example, \cite{Hijano:2015qja, Fitzpatrick:2015foa, Alkalaev:2015lca, Fitzpatrick:2015dlt, Banerjee:2016qca, Besken:2016ooo,  daCunha:2016crm}. 

\begin{figure}[tbp]
\begin{center}
\resizebox{60mm}{!}{\includegraphics{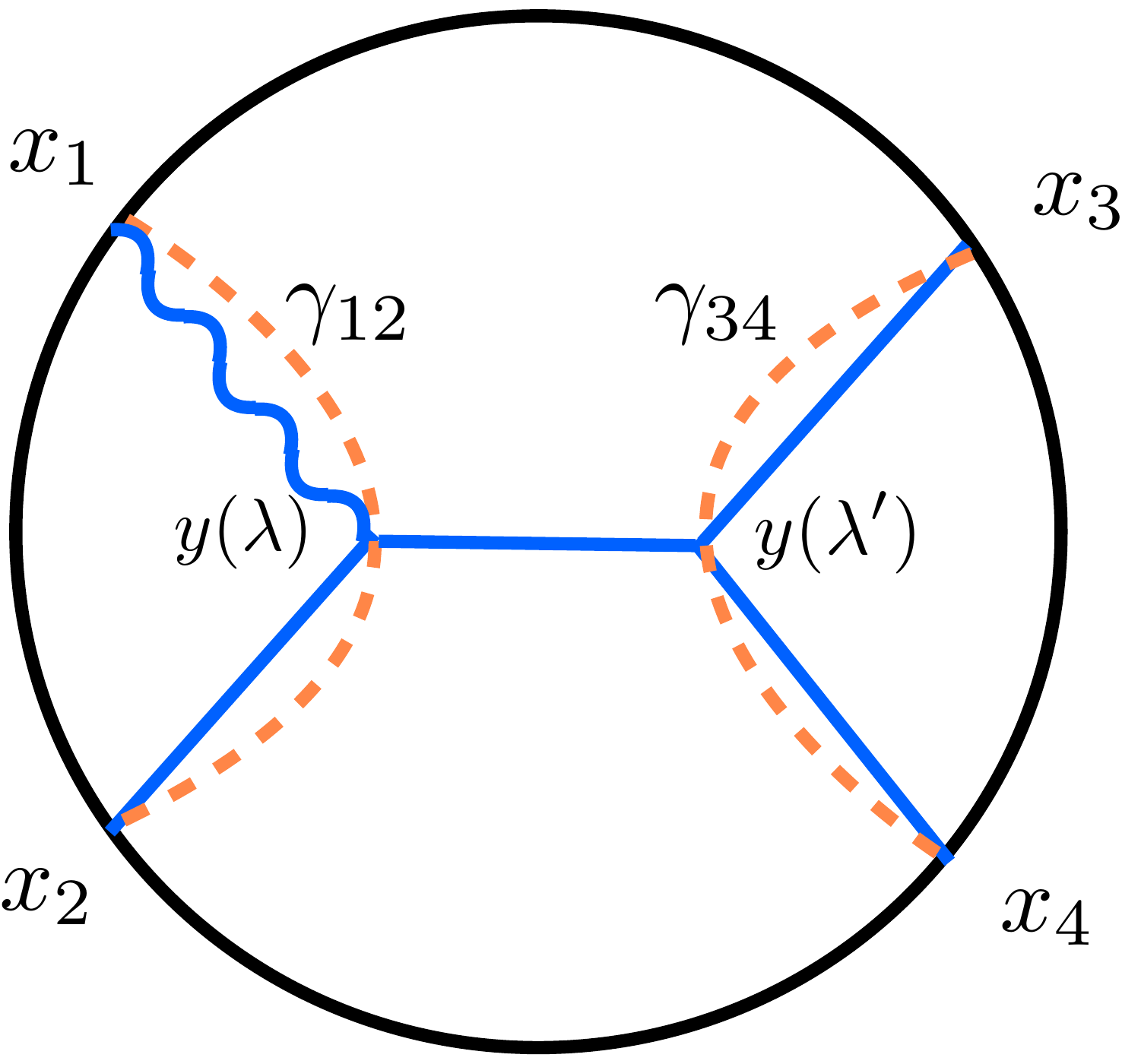}}
\caption{Scalar exchange geodesic Witten diagram with an external spin-$n$ field and three external scalar fields. The orange dashed curves are the geodesics $\gamma_{ij}$ between the boundary points ${x_i}$ and $x_{j}$. The blue wavy line represents propagation of the spin-$n$ field and the blue straight lines represent scalar propagation. The interaction vertices are integrated over the points $y$ on the geodesics $\gamma_{ij}$. The amplitude of this diagram is equivalent to the conformal partial wave up to normalization. }\label{figgwdws}
\end{center}
\end{figure}

In this paper, we extend the correspondence in \cite{Hijano:2015zsa} to the case of scalar exchange GWD of an external field with spin and three scalar fields (Figure \ref{figgwdws}). Our work is a first step toward constructing GWD with external fields in arbitrary representation. We explicitly show the correspondence (\ref{eogwd}) between CPW and GWD with an external spin-1 field up to normalization. In order to construct the amplitude of GWD, we introduce the usual three point interaction, such as $A_\mu g^{\mu\nu}\phi\partial_\nu\phi^\dagger$.  We also show that the amplitude of GWD satisfies the conformal Casimir equation. Moreover, we construct the amplitude of GWD with an external spin-$n$ field and find a three point interaction (\ref{eq:nint}) for the construction.
Our construction of GWD agrees with the known formula of CPW in \cite{Costa:2011dw}.

This paper is organized as follows. In section \ref{sec:review}, we review the correspondence between the scalar CPW and the scalar GWD in \cite{Hijano:2015zsa}. In section \ref{sec:1000}, we show the correspondence between CPW and GWD of an external spin-1 field and three external scalar fields with scalar exchange in Poincar\'{e} coordinates. In section \ref{sec:spinn}, we construct the amplitude of GWD with an external spin-$n$ field. This amplitude agrees with the known results of CPW. We also discuss three point coupling in GWD. In section \ref{sec:summary}, we summarize the results and discuss future work. We note useful formulas for our calculation in appendix \ref{app:formula} and check the relation between the scalar three point function in CFT and the amplitude of the three point scalar GWD in appendix \ref{app:3pt}. We show that GWD with an external spin-1 field satisfies the conformal Casimir equation in appendix \ref{sec:emb}.

\section{Review of conformal partial waves and geodesic Witten diagrams}\label{sec:review}
In this section, we review AdS${}_{d+1}$/CFT${}_{d}$ correspondence between conformal partial waves and geodesic Witten diagrams in \cite{Hijano:2015zsa}. We focus on scalar CPW for later analysis.

In conformal field theories, four point functions of primary operators can be expanded by CPW $W_{\Delta,\ell}(x_i)$\footnote{For simplicity, we consider the symmetric traceless representation only and we suppress the index of spin in this section. Generally, the independent number of CPW is not one if we consider nonzero external spin.} (see, for example, \cite{Rychkov:2016iqz, Simmons-Duffin:2016gjk}),
\begin{align}
\langle\mathcal{O}_1(x_1)\mathcal{O}_2(x_2)\mathcal{O}_3(x_3)\mathcal{O}_4(x_4)\rangle=\sum_{\mathcal{O}}C_{12\mathcal{O}}C^\mathcal{O}_{\;\;\;34}W_{\Delta, \ell}(x_i), \label{fpf} 
\end{align}
where $\mathcal{O}$ is a primary operator with conformal dimension $\Delta$ and spin $\ell$, $C_{12\mathcal{O}}$ and $C^\mathcal{O}_{\;\;\;34}$ are the OPE coefficients. If $\mathcal{O}_i$ are the scalar primary fields with conformal dimension $\Delta_i$, the conformal block $G_{\Delta, \ell}(u, v)$ is related to CPW\,$W_{\Delta, \ell}(x_i; \Delta_i)$,
\begin{align}
W_{\Delta, \ell}(x_i; \Delta_i)=\left(\frac{x_{24}^2}{x_{14}^2}\right)^{\frac{1}{2}\Delta_{12}}\left(\frac{x_{14}^2}{x_{13}^2}\right)^{\frac{1}{2}\Delta_{34}}\frac{G_{\Delta, \ell}(u,v)}{(x_{12}^2)^{\frac{1}{2}(\Delta_1+\Delta_2)}(x_{34}^2)^{\frac{1}{2}(\Delta_3+\Delta_4)}}, 
\end{align}
where $\Delta_{ij}\equiv\Delta_i-\Delta_j$, $x_{ij}\equiv x_i-x_j$ and $u, v$ are conformal cross ratios
\begin{align}
u=\frac{x_{12}^2x_{34}^2}{x_{13}^2x_{24}^2},\;\; v=\frac{x_{14}^2x_{23}^2}{x_{13}^2x_{24}^2}.
\end{align}
Conformal blocks and CPW can be determined from conformal symmetry and they do not depend on the details of CFT. 

\begin{figure}[tbp]
\begin{center}
\resizebox{60mm}{!}{\includegraphics{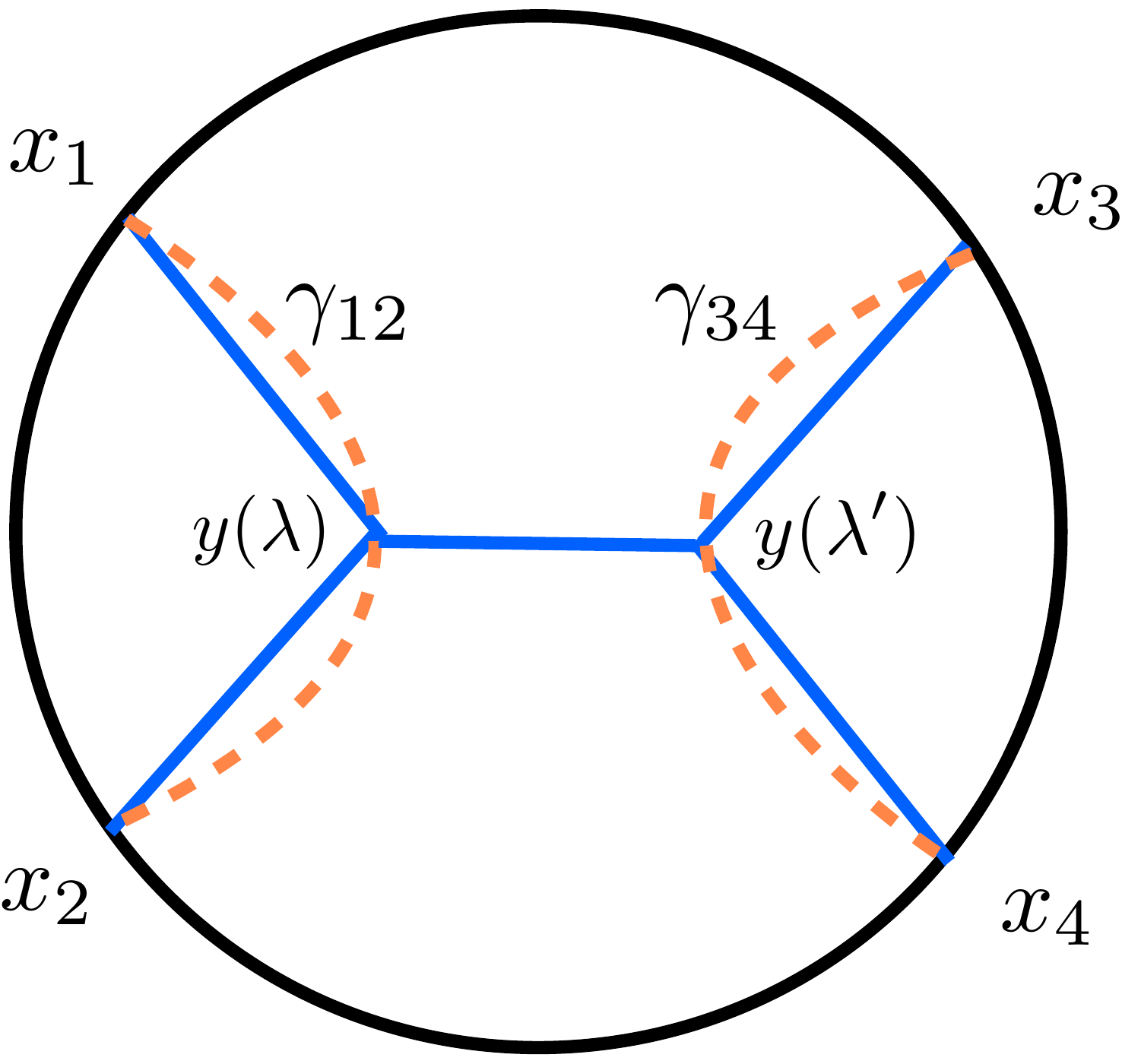}}
\caption{Scalar exchange geodesic Witten diagram with four external scalar fields. Each line has the same meaning as the lines in Figure \ref{figgwdws}. }\label{figsgwd}
\end{center}
\end{figure}

As noted above, GWD was proposed as the gravity dual of CPW up to normalization. We can define the amplitude of GWD by integrating the bulk vertices over geodesics between external fields. For example, the amplitude of the scalar exchange GWD with four external scalar fields $\mathcal{W}_{\Delta, 0}(x_i; \Delta_i)$ (Figure \ref{figsgwd}) is defined as 
\begin{align}
\mathcal{W}_{\Delta, 0}(x_i; \Delta_i)\equiv\int^\infty_{-\infty}d\lambda\int^\infty_{-\infty}d\lambda'&G_{b\partial}(y(\lambda), x_1; \Delta_1)G_{b\partial}(y(\lambda),x _2; \Delta_2)G_{bb}(y(\lambda), y(\lambda'); \Delta)\notag\\
\times &G_{b\partial}(y(\lambda'), x_3; \Delta_3)G_{b\partial}(y(\lambda'), x _4; \Delta_4)\label{sgwd}, 
\end{align}
where $\lambda$ and $\lambda'$ are proper time coordinates of geodesics $\gamma_{12}$ and $\gamma_{34}$. The terms $\gamma_{ij}$ are the geodesics between boundary points $x_i$ and $x_j$. $y(\lambda)$ and $y(\lambda')$ are coordinates of $\gamma_{12}$ and $\gamma_{34}$. The bulk-boundary propagator and the bulk-bulk propagator on AdS spacetime are denoted by $G_{b\partial}$ and $G_{bb}$, respectively\footnote{Our normalization is the convention of \cite{Hijano:2015zsa}.} (see, for example, \cite{D'Hoker:2002aw, Penedones:2016voo}), 
\begin{align}
G_{b\partial}(y, x_i; \Delta_i)&\equiv\left(\frac{u}{u^2+|x-x_i|^2}\right)^{\Delta_i}, \label{eq:Gbp}\\
G_{bb}(y, y'; \Delta)&\equiv\xi^\Delta\!_2F_1\left(\frac{\Delta}{2}, \frac{\Delta+1}{2}, \Delta+1-\frac{d}{2}; \xi^2\right),\label{eq:Gbb}\\
\xi&\equiv\frac{2uu'}{u^2+u'^2+|x-x'|^2}. \label{eq:Gbbxi}
\end{align}
Now, we consider $(d+1)$-dimensional Poincar\'e coordinates $y^\mu=\{u, x^a\}$ and the metric is \footnote{We fix the AdS radius as $R_{AdS}=1$.}
\begin{align}
ds^2=\frac{du^2+dx^adx^a}{u^2}.
\end{align}
The explicit form of $y(\lambda)$ is
\begin{align}
u(\lambda)&=\frac{|x_1-x_2|}{2\cosh{\lambda}},\label{dou}\\
x^a(\lambda)&=\frac{x^a_1+x_2^a}{2}-\frac{x^a_1-x^a_2}{2}\tanh{\lambda},\label{dox}
\end{align}
and the same is true of $y(\lambda')$.

Surprisingly, (\ref{sgwd}) is the same form of a double integral representation for the scalar CPW in \cite{Ferrara:1973vz, Ferrara:1974nf}.\footnote{In particular, \cite{Ferrara:1973vz}'s equation (32) corresponds to (\ref{sgwd}) ($d=4$) with change of variables,
\begin{align}
u=\frac{e^{-2\lambda}}{1+e^{-2\lambda}},\;\; v=\frac{e^{-2\lambda'}}{1+e^{-2\lambda'}},\;\;\lambda_{+}=\xi^{-1}. \nn
\end{align}}
Moreover, one can show that (\ref{sgwd}) satisfies the conformal Casimir equation by using embedding formalism. Therefore, we can conclude that the amplitude of GWD corresponds to CPW. Some readers may wonder why we use the exchange GWD rather than contact diagrams. This is because an equation of the bulk-bulk propagator in the scalar exchange GWD corresponds to the conformal Casimir equation and  the contact GWD do not satisfy the conformal Casimir equation.  In the next section, we will extend this result to the correspondence with an external spin-$1$ field.

\section{Direct proof of the correspondence with an external spin-$1$ field}\label{sec:1000}
In this section, we show the correspondence between conformal partial waves and geodesic Witten diagrams of an external spin-1 field and three scalar fields with scalar exchange up to normalization. We derive CPW with an external spin-1 field explicitly based on (\ref{sgwd}) and rewrite it in terms of the spin-1 propagator in AdS spacetime. 

There is a useful formula, called  the shadow formalism \cite{Ferrara:1973vz, Ferrara:1974nf,
Ferrara:1972xe, 
Ferrara:1972ay, Ferrara:1972uq} for computing CPW. We review it based on \cite{SimmonsDuffin:2012uy}. Consider a scalar primary operator $\mathcal{O}(x)$ with conformal dimension $\Delta$ and its shadow operator $\widetilde{\mathcal{O}}(x)$, which is a scalar operator with conformal dimension $d-\Delta$. Then $\widetilde{\mathcal{O}}(x)$ can be defined as
\begin{align}
\widetilde{\mathcal{O}}(x)\equiv\int d^dx'\frac{\mathcal{O}(x')}{|x'-x|^{2(d-\D)}}.
\end{align}
Let us introduce an integral
\be
\int d^dx\mathcal{O}(x)|0\rangle\langle0|\widetilde{\mathcal{O}}(x)\label{so}
\ee
and insert (\ref{so}) into (\ref{fpf}). This insertion becomes a projection into CPW,
\be
\int d^dx \langle\mathcal{O}_1(x_1)\mathcal{O}_2(x_2)\mathcal{O}(x)\rangle\langle\widetilde{\mathcal{O}}(x)\mathcal{O}_3(x_3)\mathcal{O}_4(x_4)\rangle\propto W_{\Delta, 0}(x_i)+K_{\mathcal{O}}W_{d-\Delta, 0}(x_i),\label{sf}
\ee
where $K_{\mathcal{O}}$ is a constant. This is because the l.h.s of (\ref{sf}) satisfies the conformal Casimir equation and the insertion of (\ref{so}) does not change the transformation properties. $W_{d-\Delta, 0}(x_i)$ is the shadow CPW and its boundary condition is different from $W_{\Delta, 0}(x_i)$'s boundary condition at $x_{ij}\to 0$. We can ignore $K_{\mathcal{O}}W_{d-\Delta, 0}(x_i)$ by imposing the appropriate boundary condition.

In preparation for our calculation, we consider the relation between three point functions since we integrate a product of the three point functions in the shadow formalism. The forms of the three point functions in CFT are determined by conformal symmetry\footnote{We ignore the OPE coefficients.},  
\begin{align}
\langle\mathcal{O}_1(x_1)\mathcal{O}_2(x_2)\mathcal{O}_3(x_3)\rangle=&\frac{1}{|x_{12}|^{\Delta_1+\Delta_2-\D_3}|x_{23}|^{\Delta_2+\Delta_3-\D_1}|x_{31}|^{\Delta_3+\Delta_1-\D_2}},\label{stpf}
\\
\langle\mathcal{J}^a(x_1)\mathcal{O}_2(x_2)\mathcal{O}_3(x_3)\rangle=&\frac{1}{|x_{12}|^{\Delta_1+\Delta_2-\D_3}|x_{23}|^{\Delta_2+\Delta_3-\D_1}|x_{31}|^{\Delta_3+\Delta_1-\D_2}}\times\left(\frac{x_{12}^a}{|x_{12}|^2}-\frac{x_{13}^a}{|x_{13}|^2}\right), \label{vtpf}
\end{align}
where $\mathcal{O}_i(x_i)$ are scalar primary fields with conformal dimension $\D_i$ and $\mathcal{J}^a(x_1)$ is a spin-1 primary field with conformal dimension $\D_1+1$.  The relation between (\ref{stpf}) and (\ref{vtpf}) is
\be
\left(\frac{\partial}{\partial x^a_1}+\frac{2\D_1(x_{12})_a}{|x_{12}|^2}\right)\langle\mathcal{O}_1(x_1)\mathcal{O}_2(x_2)\mathcal{O}_3(x_3)\rangle=(\D_3+\D_1-\D_2)\langle\mathcal{J}_a(x_1)\mathcal{O}_2(x_2)\mathcal{O}_3(x_3)\rangle.\label{rtps}
\ee

For the remainder of the paper, we denote CPW of four external primary fields with conformal dimension $\D_i$ and spin $\ell_i$ by  $W^{(\ell_1, \ell_2, \ell_3, \ell_4)}_{\D, \ell}(x_i; \D_i)$\footnote{The independent number of CPW that we consider in this paper is one. This is because the degrees of freedom of the three point functions that we consider are one such as (\ref{stpf}) and (\ref{vtpf}).}. Here $\D$ and $\ell$ are the conformal dimension and spin of an exchanging primary operator.  Similarly, we denote the amplitude of GWD as $\mathcal{W}^{(\ell_1, \ell_2, \ell_3, \ell_4)}_{\D, \ell}(x_i; \D_i)$.
By using  (\ref{sf}) and (\ref{rtps}), we get (up to normalization)
\begin{align}
\left(W^{(1, 0, 0, 0)}_{\D, 0}(x_i; \widetilde{\D}_i)\right)_a&=\int d^dx \langle\mathcal{J}_a(x_1)\mathcal{O}_2(x_2)\mathcal{O}(x)\rangle\langle\widetilde{\mathcal{O}}(x)\mathcal{O}_3(x_3)\mathcal{O}_4(x_4)\rangle|_{\textrm{BC }}\notag\\
&=\left(\frac{\partial}{\partial x^a_1}+\frac{2\D_1(x_{12})_a}{|x_{12}|^2}\right)\int d^dx \langle\mathcal{O}_1(x_1)\mathcal{O}_2(x_2)\mathcal{O}(x)\rangle\langle\widetilde{\mathcal{O}}(x)\mathcal{O}_3(x_3)\mathcal{O}_4(x_4)\rangle|_{\textrm{BC}}\notag\\
&=\left(\frac{\partial}{\partial x^a_1}+\frac{2\D_1(x_{12})_a}{|x_{12}|^2}\right)W^{(0, 0, 0, 0)}_{\D, 0}(x_i; \D_i)\notag\\
&=\left(\frac{\partial}{\partial x^a_1}+\frac{2\D_1(x_{12})_a}{|x_{12}|^2}\right)\mathcal{W}^{(0, 0, 0, 0)}_{\D, 0}(x_i; \D_i), \label{efcpw}
\end{align}
where $\widetilde{\D}_i=\D_i+\delta_{i1}$. 
Here $|_{\textrm{BC}}$ means imposing the appropriate boundary condition to ignore the shadow CPW and the explicit forms of the boundary conditions for CPW are
\begin{align}
\lim_{x_{12}\to0}W^{(0, 0, 0, 0)}_{\D, 0}(x_i; \D_i)&\to\frac{(\textrm{constant})}{|x_{12}|^{\D_1+\D_2-\D}},\\
\lim_{x_{12}\to0}\left(W^{(1, 0, 0, 0)}_{\D, 0}(x_i; \widetilde{\D}_i)\right)_a&\to\frac{(\textrm{constant})}{|x_{12}|^{\D_1+\D_2-\D}}\times\frac{(x_{12})_a}{|x_{12}|^2}.
\end{align}
Thus, we have obtained formula (\ref{efcpw}) of CPW with an external spin-1 field $W^{(1, 0, 0, 0)}_{\D, 0}(x_i; \widetilde{\D}_i)$ in terms of (\ref{sgwd}).

However, the relationship between (\ref{efcpw}) and the spin-1 propagator in AdS spacetime is not clear. In order to make it manifest, we rewrite (\ref{efcpw}) in terms of the spin-1 propagator. In particular, we will show\footnote{We note that $\partial/\partial x^a_1$ acts on both $x_1$ and $y(\lambda)$.}
\begin{align}
\left(\frac{\partial}{\partial x^a_1}+2\D_1\frac{(x_{12})_a}{|x_{12}|^2}\right)\int^\infty_{-\infty}d\lambda\int^\infty_{-\infty}d\lambda'&G_{b\partial}(y(\lambda), x_1; \Delta_1)G_{b\partial}(y(\lambda),x _2; \Delta_2)G_{bb}(y(\lambda), y(\lambda'); \Delta)\notag\\
\times &G_{b\partial}(y(\lambda'), x_3; \Delta_3)G_{b\partial}(y(\lambda'), x _4; \Delta_4)\notag\\
=\int^\infty_{-\infty}d\lambda\int^\infty_{-\infty}d\lambda'&\left(G^1_{b\partial}(y(\lambda), x_1; \Delta_1+1)\right)_{a}^{\mu}G_{b\partial}(y(\lambda),x _2; \Delta_2)\notag\\
\times& u(\lambda)^2\frac{\partial}{\partial y^{\mu}(\lambda)}\left(G_{bb}(y(\lambda), y(\lambda'); \Delta)\right)\notag\\
\times &G_{b\partial}(y(\lambda'), x_3; \Delta_3)G_{b\partial}(y(\lambda'), x _4; \Delta_4),
\label{eogwd}\end{align}
where $G^1_{b\partial}(y, x_1; \Delta_1+1)$ is the spin-1 bulk-boundary propagator (see, for example, \cite{Freedman:1998tz, Costa:2014kfa}),
\begin{align}
\left(G^1_{b\partial}(y, x_1; \Delta_1+1)\right)_{a}^{\mu}\equiv\left(\frac{u}{u^2+|x-x_1|^2}\right)^{\Delta_1}\left(\frac{\delta_{a}^{\mu}}{u^2+|x-x_1|^2}-2\frac{(y-x_1)_a(y-x_1)^\mu}{(u^2+|x-x_1|^2)^2}\right). \label{eq:Gbp1}
\end{align}
The l.h.s of (\ref{eogwd}) corresponds to the last line of (\ref{efcpw}). The r.h.s of (\ref{eogwd}) is a definition of the amplitude of GWD $\mathcal{W}^{(1, 0, 0, 0)}_{\D, 0}(x_i; \widetilde{\D}_i)$ with a three point interaction coefficient $u^2\frac{\partial}{\partial y^{\mu}}$ that is the usual coupling such as $A_\mu g^{\mu\nu}\phi\partial_\nu\phi^\dagger$\footnote{Note that this interaction is one of the candidates. One can obtain the same tensor structure by using $A_\mu g^{\mu\nu}\nabla^2\phi\partial_\nu\phi^\dagger$, for example. Such three point interactions must be invariant under the isometry of AdS for the correspondence between CPW and GWD. In contrast to the Witten diagrams, forms of the three point interactions in GWD have no physical meaning because CPW does not depend on the dynamics.}. Therefore, (\ref{eogwd}) signifies the correspondence between CPW and GWD with an external spin-1 field.

In order to show (\ref{eogwd}), we deform
\begin{align}
\frac{\partial}{\partial x^a_1}\int^\infty_{-\infty}d\lambda&G_{b\partial}(y(\lambda), x_1; \Delta_1)G_{b\partial}(y(\lambda),x _2; \Delta_2)G_{bb}(y(\lambda), y(\lambda'); \Delta).
\end{align}
From (\ref{dou}), (\ref{dox}) and the definitions of the propagators, we find
\begin{align}
&\frac{\partial}{\partial x^a_1}\int^\infty_{-\infty}d\lambda G_{b\partial}(y(\lambda), x_1; \Delta_1)G_{b\partial}(y(\lambda),x _2; \Delta_2)G_{bb}(y(\lambda), y(\lambda'); \Delta)\notag\\
=&-(\D_1+\D_2)\frac{(x_{12})_a}{|x_{12}|^2}\int^\infty_{-\infty}d\lambda G_{b\partial}(y(\lambda), x_1; \Delta_1)G_{b\partial}(y(\lambda),x _2; \Delta_2)G_{bb}(y(\lambda), y(\lambda'); \Delta)\notag\\
&+\int^\infty_{-\infty}d\lambda\left(G^1_{b\partial}(y(\lambda), x_1; \Delta_1+1)\right)_{a}^{\mu}G_{b\partial}(y(\lambda),x _2; \Delta_2)u(\lambda)^2\frac{\partial}{\partial y^{\mu}(\lambda)}\left(G_{bb}(y(\lambda), y(\lambda'); \Delta)\right)\notag\\
&-\frac{(x_{12})_a}{|x_{12}|^2}\int^\infty_{-\infty}d\lambda G_{b\partial}(y(\lambda), x_1; \Delta_1)G_{b\partial}(y(\lambda),x _2; \Delta_2)\frac{\p y^\mu(\lambda)}{\p \lambda}\frac{\partial}{\partial y^\mu(\lambda)}G_{bb}(y(\lambda), y(\lambda'); \Delta).\label{e310}
\end{align}
After integration by parts of the last line of (\ref{e310}), we obtain\footnote{We assume $|\D_1-\D_2|<\D$ for integration by parts. This is the same condition for the convergence of the amplitude of GWD for the scalar three point function. We note the correspondence between the scalar three point function and GWD in appendix \ref{app:3pt}.}
\begin{align}
&\frac{\partial}{\partial x^a_1}\int^\infty_{-\infty}d\lambda G_{b\partial}(y(\lambda), x_1; \Delta_1)G_{b\partial}(y(\lambda),x _2; \Delta_2)G_{bb}(y(\lambda), y(\lambda'); \Delta)\notag\\
=&-2\D_1\frac{(x_{12})_a}{|x_{12}|^2}\int^\infty_{-\infty}d\lambda G_{b\partial}(y(\lambda), x_1; \Delta_1)G_{b\partial}(y(\lambda),x _2; \Delta_2)G_{bb}(y(\lambda), y(\lambda'); \Delta)\notag\\
&+\int^\infty_{-\infty}d\lambda\left(G^1_{b\partial}(y(\lambda), x_1; \Delta_1+1)\right)_{a}^{\mu}G_{b\partial}(y(\lambda),x _2; \Delta_2)u(\lambda)^2\frac{\partial}{\partial y^{\mu}(\lambda)}\left(G_{bb}(y(\lambda), y(\lambda'); \Delta)\right).\label{e311}
\end{align}
Integrating (\ref{e311}) by $\lambda'$ with $G_{b\partial}(y(\lambda'), x_3; \Delta_3)G_{b\partial}(y(\lambda'), x _4; \Delta_4)$, we obtain the final result (\ref{eogwd}).  The formulas in appendix \ref{app:formula} are useful for the calculation.

Summarizing the above, we have explicitly shown the correspondence (\ref{eogwd}) between the conformal partial wave $W^{(1, 0, 0, 0)}_{\D, 0}(x_i; \widetilde{\D}_i)$ and the amplitude of the geodesic Witten diagram $\mathcal{W}^{(1, 0, 0, 0)}_{\D, 0}(x_i; \widetilde{\D}_i)$ of an external spin-1 field and three scalar fields with scalar exchange. We can prove this correspondence by using conformal Casimir equation (see appendix \ref{sec:emb}).

\section{Generalization to an external spin-$n$ field}\label{sec:spinn}
In this section, we extend the previous result to the geodesic Witten diagrams with an external spin-$n$ field. To see the equivalence between GWD and CPW, it is useful to employ so-called embedding formalism. In section \ref{subsec:emb}, we review the embedding formalism in order to note our notation. In section \ref{subsec:spinn}, we specify three point coupling in GWD and explicitly construct the amplitude of GWD with an external spin-$n$ field $\mathcal{W}^{(n, 0, 0, 0)}_{\D, 0}(x_i; \D_i)$. This expression agrees with the formula of CPW in \cite{Costa:2011dw}. 
\subsection{Embedding space}\label{subsec:emb}
It is a well-known fact that the conformal symmetry of $d$-dimensional CFT and the isometry of Euclidean AdS${}_{d+1}$ are equivalent to ($d+2$)-dimensional Lorentz symmetry. By using this fact, we can describe CFT in $d$-dimension and a theory on AdS${}_{d+1}$ space as a theory on $d+2$-dimensional embedding Minkowski spacetime.
This formalism is called embedding formalism. (see, for example, \cite{Dirac:1936fq,Boulware:1970ty,Ferrara:1973yt,Ferrara:1973eg,Cornalba:2009ax,Weinberg:2010fx,Costa:2011mg, Costa:2011dw,Costa:2014kfa}.) 
Since the Lorentz transformation is linear, tensor structures in the embedding formalism become simple.
From the above motivation, we review the embedding formalism. For more details about the embedding formalism in CFT, see \cite{Costa:2011mg, Costa:2011dw}. One can find the details of the embedding formalism for AdS in \cite{Costa:2014kfa, Bekaert:2015tva}. 

\begin{figure}[tbp]
\begin{center}
\resizebox{80mm}{!}{\includegraphics{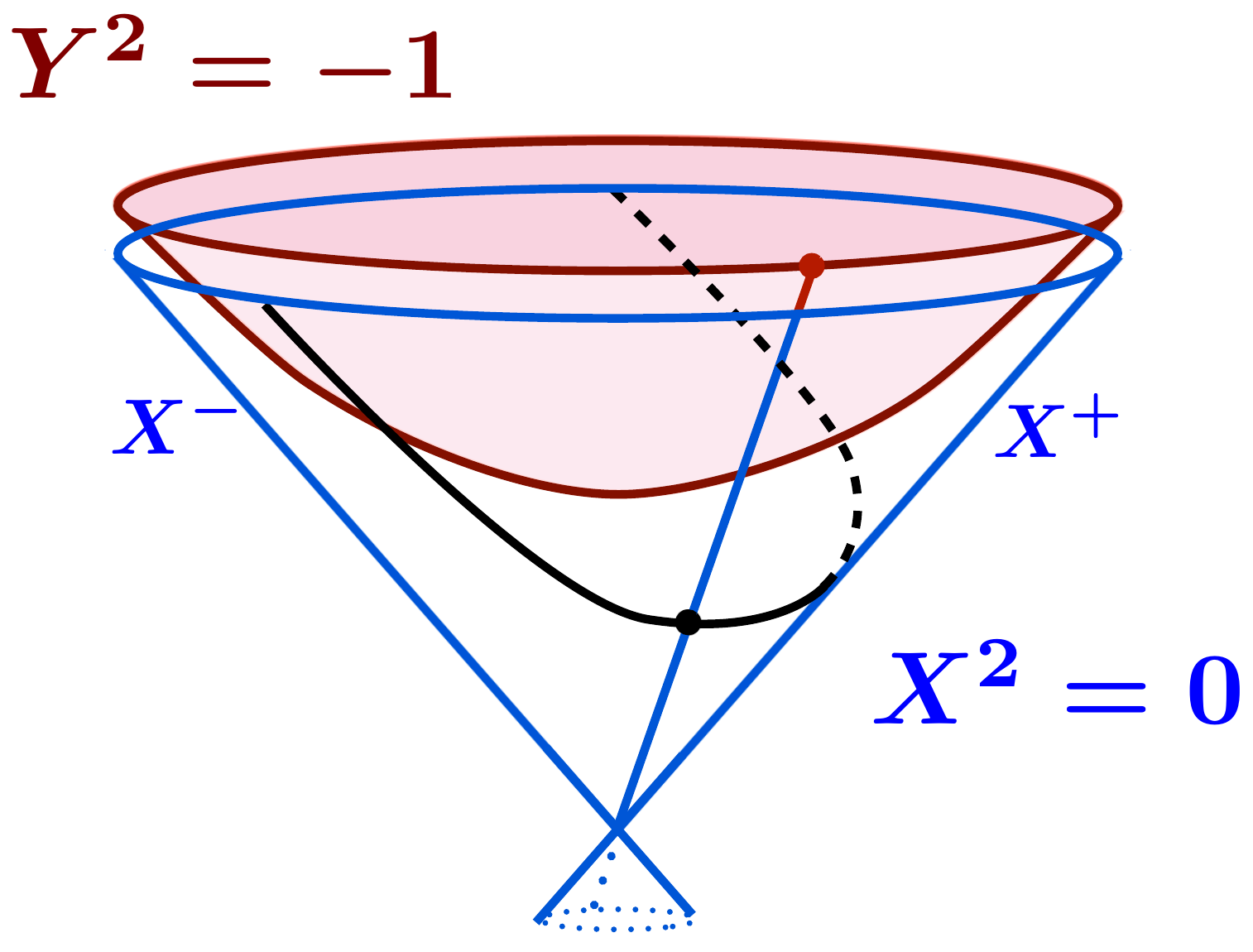}}
\caption{Euclidean AdS (red hyperboloid) and its conformal boundary (blue light cone) in the embedding space. The blue light ray shows the identification of the boundary points $X^A\sim\la X^A$. The black hyperbolic curve displays one choice of the flat section for CFT (the Poincar\'e section). }\label{fig:nullandads}
\end{center}
\end{figure} 

Euclidean AdS${}_{d+1}$ can be embedded into ($d+2$)-dimensional Minkowski spacetime $\mathbb{R}^{1,d+1}$ as
\be
Y^2=\eta_{AB}Y^AY^B=-1,\; Y^0>0, 
\ee
where $Y^A$ denotes the coordinates of $\mathbb{R}^{1,d+1}$. 
We embed AdS${}_{d+1}$ coordinates $y^\mu=\{u, x^a\}$ into $Y^A$ such that
\begin{align}
Y^A&\equiv(Y^+, Y^-, Y^a) \\
&=\dfrac{1}{u}(1,u^2+x^2,x^a).
\end{align}
The conformal boundary of AdS, on which CFT lives, can be defined as the projective light cone
\be 
X^2=0,\; X^A\sim \la X^A \hspace{5mm}(\la\in\mathbb{R}) . 
\ee
We use the Poincar\'e section (Figure \ref{fig:nullandads}) for the $d$-dimensional flat space $\mathbb{R}^d$, 
\begin{align}
X^A&\equiv(X^+, X^-, X^a) \nn\\
&=(1,x^2,x^a).
\end{align}

Next, we embed the fields in CFT${}_{d}$ and AdS${}_{d+1}$ into the embedding space. Since the number of the fields in the embedding space is larger than that of the fields in CFT and AdS, we must impose the constraints for the fields in the embedding space. 
In particular, we impose a transverse condition to traceless symmetric tensors in both sides as
\be
X_{A_1}T_{\p}^{A_1A_2\cdots A_l}(X)=0,\hspace{3mm} Y_{A_1}T_{b}^{A_1A_2\cdots A_l}(Y)=0, \label{eq:transverse}
\ee
where $T_{\p}$ is the tensor field in the boundary CFT and $T_{b}$ is in the bulk AdS. 
We further impose the condition to the primary field $T_{\p}^{A_1A_2\cdots A_l}(X)$ as
\be
T_{\p}^{A_1A_2\cdots A_l}(\la X)=\la^{-\D}T_{\p}^{A_1A_2\cdots A_l}(X). 
\ee
When we consider the tensor fields, there is an efficient way to classify their tensor structures, which are called index-free notation, introduced in \cite{Costa:2011mg, Costa:2011dw, Costa:2014kfa}. 
For this notation, we introduce auxiliary fields $Z$ for the boundary, and $W$ for the bulk to contract all indices:
\be
T_{\p}(X; Z)\equiv Z_{A_1}\cdots Z_{A_l}T_{\p}^{A_1A_2\cdots A_l}(X),\hspace{3mm}T_{b}(Y; W)\equiv W_{A_1}\cdots W_{A_l}T_{b}^{A_1A_2\cdots A_l}(Y). 
\ee
We can restrict $Z$ to $Z^{2}=Z \cdot X=0$ because these conditions do not lose the information of $T_{\p}$. Similarly, we can restrict $W$ to $W^2=W\cdot Y=0$. 

With index-free notation, we define the bulk-boundary propagators in embedding space \cite{Costa:2014kfa} as
\be
G^J_{b\p}(X, Y; Z, W; \D)\equiv\dfrac{\left((-2X\cdot Y)(Z\cdot W)+2(Z\cdot Y)(X\cdot W)\right)^J}{(-2X\cdot Y)^{\D+J}}. \label{eq:embbdry}
\ee 
One can remove the auxiliary fields $Z$ and $W$ by using some differential operators introduced in \cite{Costa:2011mg, Costa:2011dw, Costa:2014kfa}. 
Since the numerator of \eqref{eq:embbdry} is written as $J$ copies of the first-order polynomial of $Z$ (or $W$), to remove $Z$ and $W$ it is enough to use $\frac{\p}{\p Z}$ and $\frac{\p}{\p W}$ naively.
One can also see that the bulk-bulk propagator in the embedding space is
\begin{align}
G_{bb}(Y_1, Y_2; \Delta)&\equiv\xi^\Delta\!_2F_1\left(\frac{\Delta}{2}, \frac{\Delta+1}{2}, \Delta+1-\frac{d}{2}; \xi^2\right),\\
\xi^{-1}&\equiv-Y_1\cdot Y_2. \label{eq:xiemb}
\end{align}
Here \eqref{eq:xiemb} is the expression of \eqref{eq:Gbbxi} in the embedding formalism.  

By employing the above ingredients, we can rewrite the amplitude of the scalar GWD \eqref{sgwd} as following: 
\begin{align}
\mathcal{W}^{(0,0,0,0)}_{\D,0}(X_i; \D_i)&=\int^{\infty}_{-\infty} d\la^\prime\,\left[\int^{\infty}_{-\infty} d\la\, G^0_{b\p}(Y_1(\la),X_1, \D_1)\,G^0_{b\p}(Y_1(\la),X_2; \D_2)\,G_{bb}(Y_1(\la),Y_2(\la^\prime); \D)\right] \nn\\
&\times G^0_{b\p}(Y_2(\la^\prime),X_3; \D_3)\,G^0_{b\p}(Y_2(\la^\prime),X_4; \D_4). 
\end{align}
Here $Y_{1A}(\la)$, which is the point on the geodesic between the boundary points $X_1$ and $X_2$, can be written simply as
\be
Y_{1A}(\la)=\dfrac{e^{-\la}X_{1A}+e^{\la}X_{2A}}{\sqrt{-2X_1\cdot X_2}}.
\ee
Similarly, $Y_{2}(\la^\prime)$ can be written as
\be
Y_{2A}(\la^\prime)=\dfrac{e^{-\la^\prime}X_{3A}+e^{\la^\prime}X_{4A}}{\sqrt{-2X_3\cdot X_4}}. 
\ee
On the other hand, the amplitude of GWD $\mathcal{W}^{(1,0,0,0)}_{\D,0}(X_i; Z_1; \D_i)$ can be rewritten in the index-free notation as
\begin{align}
&\hspace{-1mm}\mathcal{W}^{(1,0,0,0)}_{\D,0}(X_i; Z_1; \D_i)\nn\\
&=\int^{\infty}_{-\infty} d\la^\prime\,\left[\int^\infty_{-\infty} d\la\,G^0_{b\p}(Y_1(\la),X_2; \D_2)\left\{G^1_{b\p}(Y_1(\la),X_1;Z_1,\nabla_{Y_1}; \D_1)\,G_{bb}(Y_1(\la),Y_2(\la^\prime); \D)\right\}\right]\nn\\
&\times G^0_{b\p}(Y_2(\la^\prime),X_3; \D_3)\,G^0_{b\p}(Y_2(\la^\prime),X_4; \D_4). \label{eq:efcpw}
\end{align}
Here we introduce a covariant derivative in the embedding AdS space \cite{Costa:2014kfa} as
\be
\nabla_A\equiv\dfrac{\p}{\p Y^A}+Y_A\left(Y\cdot\dfrac{\p}{\p Y}\right)+W_A\left(Y\cdot\dfrac{\p}{\p W}\right),
\ee
which also satisfies $Y^A\nabla_A=0$. 
By using formulas in appendix \ref{app:formula}, one can check that \eqref{eq:efcpw} is equivalent to the r.h.s of \eqref{eogwd}. 

One can translate the previous result \eqref{eogwd} into
\be
\left(Z_1\cdot\dfrac{\p}{\p X_1}-2\D_1\dfrac{Z_1\cdot X_2}{(-2X_1\cdot X_2)}\right)\mathcal{W}^{(0,0,0,0)}_{\D,0}(X_i; \D_i)=\mathcal{W}^{(1,0,0,0)}_{\D,0}(X_i;Z_1;\widetilde{\D}_i), \label{eq:1and0}
\ee
where $\widetilde{\D}_i\equiv\D_i+\delta_{i1}$.
\eqref{eq:1and0} represents the relation between the scalar GWD and GWD with an external spin-1 field in the embedding formalism. 
Of course, one can check this relation directly from \eqref{eq:efcpw}. 
For the explicit computation to check \eqref{eq:1and0} in the embedding formalism, appendix \ref{app:formula} may be useful. 
For later convenience, we define
\begin{align}
F^{(0,0; 0)}_{\D_1,\D_2,\D}(X_1,X_2,Y_2)&\equiv\int^{\infty}_{-\infty} d\la\, G^0_{b\p}(Y_1,X_1; \D_1)\,G^0_{b\p}(Y_1,X_2; \D_1)\,G_{bb}(Y_1,Y_2; \D), \nn\\
F^{(n,0; 0)}_{\D_1,\D_2,\D}(X_1,X_2,Y_2;Z_1)&\equiv\int^\infty_{-\infty} d\la\,G^0_{b\p}(Y_1,X_2; \D_2)\left\{G^n_{b\p}(Y_1,X_1;Z_1,\nabla_{Y_1}; \D_1)\,G_{bb}(Y_1,Y_2; \D)\right\}, \nn\\
F^{(0,n;0)}_{\D_1,\D_2,\D}(X_1,X_2,Y_2;Z_2)&\equiv\int^\infty_{-\infty} d\la\,G^0_{b\p}(Y_1,X_1; \D_1)\left\{G^n_{b\p}(Y_1,X_2;Z_2,\nabla_{Y_1}; \D_2)\,G_{bb}(Y_1,Y_2; \D)\right\}. 
\end{align}
Here, $G^n_{b\p}(Y,X;Z,\nabla; \D)$ denotes
\be
\frac{1}{(-2X\cdot Y)^{\D+n}}\underbrace{\left[(-2X\cdot Y)(Z\cdot \nabla)+2(Z\cdot Y)(X\cdot \nabla)\right]\cdots\left[(-2X\cdot Y)(Z\cdot \nabla)+2(Z\cdot Y)(X\cdot \nabla)\right]}_{\;\large n\;}. 
\ee 
\subsection{Explicit construction for an external spin-$n$ field}\label{subsec:spinn}
Towards the generalization to arbitrary symmetric-traceless representation, we construct the amplitude of GWD with an external spin-$n$ field.
We also show that our construction of GWD agrees with the expression in \cite{Costa:2011dw}. Moreover, we discuss a corresponding three point interaction in the bulk.

Based on \eqref{eq:efcpw}, it is straightforward to define GWD $\mathcal{W}^{(n,0,0,0)}_{\D,0}$ as 
\begin{align}
&\hspace{-1mm}\mathcal{W}^{(n,0,0,0)}_{\D,0}(X_i; Z_1; \D_i)\nn\\
&\equiv\int^{\infty}_{-\infty} d\la^\prime\,\left[\int^\infty_{-\infty} d\la\,G^0_{b\p}(Y_1(\la),X_2; \D_2)\left\{G^n_{b\p}(Y_1(\la),X_1;Z_1,\nabla_{Y_1}; \D_1)\,G_{bb}(Y_1(\la),Y_2(\la^\prime); \D)\right\}\right]\nn\\
&\hspace{3cm}\times G^0_{b\p}(Y_2(\la^\prime),X_3; \D_3)\,G^0_{b\p}(Y_2(\la^\prime),X_4; \D_4) \label{eq:efcpwn}\\
&=\int^{\infty}_{-\infty} d\la^\prime\,F^{(n,0; 0)}_{\D_1,\D_2,\D}(X_1,X_2,Y_2;Z_1)G^0_{b\p}(Y_2(\la^\prime),X_3; \D_3)\,G^0_{b\p}(Y_2(\la^\prime),X_4; \D_4). 
\end{align}
In order to check that our construction agrees with the known results of CPW, it is better to understand the relation between $\mathcal{W}^{(n,0,0,0)}_{\D,0}$ and $\mathcal{W}^{(0,0,0,0)}_{\D,0}$.
To this end, it is enough to study $F^{(n,0; 0)}_{\D_1,\D_2,\D}$. 

It is known that CPW with symmetric-traceless tensors can be expressed by the scalar CPW with the differential operators such as (3.40) of \cite{Costa:2011dw}. 
They introduced the following differential operators\footnote{The authors of \cite{Costa:2011dw} used $P$ as the embedding space coordinates instead of $X$. }:
\begin{align}
\hspace{-5mm}D_{11}&\equiv(X_1\cdot X_2)\left(Z_1\cdot\dfrac{\p}{\p X_2}\right)-(Z_1\cdot X_2)\left(X_1\cdot\dfrac{\p}{\p X_2}\right)-(Z_1\cdot Z_2)\left(X_1\cdot\dfrac{\p}{\p Z_2}\right)+(X_1\cdot Z_2)\left(Z_1\cdot\dfrac{\p}{\p Z_2}\right), \nn\\
\hspace{-5mm}D_{22}&\equiv(X_2\cdot X_1)\left(Z_2\cdot\dfrac{\p}{\p X_1}\right)-(Z_2\cdot X_1)\left(X_2\cdot\dfrac{\p}{\p X_1}\right)-(Z_2\cdot Z_1)\left(X_2\cdot\dfrac{\p}{\p Z_1}\right)+(X_2\cdot Z_1)\left(Z_2\cdot\dfrac{\p}{\p Z_1}\right),\nn\\
\hspace{-5mm}D_{12}&\equiv(X_1\cdot X_2)\left(Z_1\cdot\dfrac{\p}{\p X_1}\right)-(Z_1\cdot X_2)\left(X_1\cdot\dfrac{\p}{\p X_1}\right)+(Z_1\cdot X_2)\left(Z_1\cdot\dfrac{\p}{\p Z_1}\right), \nn\\
\hspace{-5mm}D_{21}&\equiv(X_2\cdot X_1)\left(Z_2\cdot\dfrac{\p}{\p X_2}\right)-(Z_2\cdot X_1)\left(X_2\cdot\dfrac{\p}{\p X_2}\right)+(Z_2\cdot X_1)\left(Z_2\cdot\dfrac{\p}{\p Z_2}\right).
\end{align}
By using these differential operators, one can show, for example,
\begin{align}
D_{11}F^{(0,0; 0)}_{\D_1+1,\D_2,\D}&=-\dfrac{1}{2}\int^\infty_{-\infty} d\la\,G^0_{b\p}(Y_1,X_2; \D_2)\{G^1_{b\p}(Y_1,X_1;Z_1,\nabla_{Y_1}; \D_1)\,G_{bb}(Y_1,Y_2; \D)\} \nn\\
&=-\dfrac{1}{2}F^{(1,0; 0)}_{\D_1,\D_2,\D}.
\end{align}
One can repeat such manipulations and obtain
\begin{align}
(D_{11})^nF^{(0,0; 0)}_{\D_1+n,\D_2,\D}&=\left(-\dfrac{1}{2}\right)^nF^{(n,0; 0)}_{\D_1,\D_2,\D}, \label{eq:D11}\\
(D_{22})^nF^{(0,0; 0)}_{\D_1,\D_2+n,\D}&=\left(-\dfrac{1}{2}\right)^nF^{(0,n;0)}_{\D_1,\D_2,\D}, \label{eq:D22}\\
(D_{12})^nF^{(0,0; 0)}_{\D_1,\D_2+n,\D}&=\left(-\dfrac{1}{2}\right)^nF^{(n,0; 0)}_{\D_1,\D_2,\D}, \label{eq:D12}\\
(D_{21})^nF^{(0,0; 0)}_{\D_1+n,\D_2,\D}&=\left(-\dfrac{1}{2}\right)^nF^{(0,n;0)}_{\D_1,\D_2,\D}. \label{eq:D21} 
\end{align}
Now the meaning of $D_{ij}$ is clear. Namely, the action of $D_{ij}$ increases the spin of the field at $X_i$ by one and decreases the scaling dimension of the field at $X_j$ by one.
In particular, \eqref{eq:D11} and \eqref{eq:D22} agree with (3.40) of \cite{Costa:2011dw} in the case of CPW with an external spin-$n$ field and three external scalar fields. This result implies GWD can represent CPW with external tensor fields. 

Finally, we discuss a three point interaction for $\mathcal{W}^{(n,0,0,0)}_{\D,0}$. One possible answer in the embedding space is 
\be
S_{int}=\int_{\textrm{AdS}} dY\, T^{A_1\cdots A_n}_{\D_1}\phi_{\D_2}\,(\nabla_{A_1}\cdots\nabla_{A_n}\phi_{\D}). \label{eq:nint}
\ee
The reason is as follows. After simple calculation, \eqref{eq:efcpwn} becomes
\begin{align}
&\hspace{-1mm}\mathcal{W}^{(n,0,0,0)}_{\D,0}(X_i; Z_1; \D_i)\nn\\
&=\int^{\infty}_{-\infty} d\la^\prime\,\left[\int^\infty_{-\infty} d\la\,G^0_{b\p}(Y_1(\la),X_2; \D_2)\right.\nn\\
&\hspace{3cm}\times \left.G^n_{b\p}(Y_1(\la),X_1;Z_1;\D_1)^{A_1\cdots A_n}\,\dfrac{\p}{\p Y^{A_1}_1}\cdots\dfrac{\p}{\p Y^{A_n}_1}\,G_{bb}(Y_1(\la),Y_2(\la^\prime); \D)\right]\nn\\
&\hspace{3cm}\times G^0_{b\p}(Y_2(\la^\prime),X_3; \D_3)\,G^0_{b\p}(Y_2(\la^\prime),X_4; \D_4). \label{wn000}
\end{align}
Here we define
\be
G^n_{b\p}(Y,X;Z;\D)_{A_1\cdots A_n}\equiv\dfrac{1}{n!}\dfrac{\p}{\p W^{A_1}}\cdots \dfrac{\p}{\p W^{A_n}}G^n_{b\p}(Y,X;Z,W; \D). 
\ee
From (\ref{wn000}), we can choose a three point interaction to construct the amplitude of GWD $\mathcal{W}^{(n,0,0,0)}_{\D,0}$ as
\be
S_{int}=\int_{\textrm{AdS}} dY\, T^{A_1\cdots A_n}_{\D_1}\phi_{\D_2}\,\left(\dfrac{\p}{\p Y^{A_1}}\cdots\dfrac{\p}{\p Y^{A_n}}\phi_{\D}\right).
\ee
By virtue of the transverse condition \eqref{eq:transverse} and the traceless condition of $T^{A_1\cdots A_n}$, one can freely replace all of $\frac{\p}{\p Y^A}$ with $\nabla_{A}$; thus, we obtain the manifestly covariant expression \eqref{eq:nint}. 
We stress that this three point interaction is not unique one for CPW with an external spinning field.  We expect that any other three point interactions that are invariant under the isometry of AdS will give us the same tensor structure. This is because CPW in our case has unique tensor structure \cite{Costa:2011dw}.
\if(\subsection{Decomposition of Witten diagram into conformal partial waves}\label{subsec:decomp}
So far we discussed the correspondence between GWD and CPW with an external spinning field. In this subsection, we discuss the decomposition of usual tree-level Witten diagram with an external spinning field into GWD discussed above. We will see that differential operator method again relates Witten diagram with four external scalars to Witten diagram with external three scalars and one spinning field. 

We will focus on the amplitude of scalar-exchange Witten diagram: 
\begin{align}
&\hspace{-1mm}\mathcal{A}^{\textrm{Exct}}_4(\D_i;\D)\nn\\
&\equiv\int_{\textrm{AdS}} dY_1\,dY_2\, G^{0}_{b\p}(X_1,Y_1;\D_1)G^{0}_{b\p}(X_2,Y_1;\D_2)G_{bb}(Y_1,Y_2;\D)G^{0}_{b\p}(X_3,Y_2;\D_3)G^{0}_{b\p}(X_4,Y_2;\D_4). \label{eq:scalarexct}
\end{align}
We implicitly assume that all external operators $\mathcal{O}_i$ with scaling dimensions $\D_i$ are single-trace operators. 
In \cite{Hijano:2015zsa}, they showed that \eqref{eq:scalarexct} can be decomposed into the sum of scalar GWD. This decomposition can be written as following: 
\be
\mathcal{A}^{\textrm{Exch}}_4(\D_i;\D)=\alpha^{12}_{\D}\alpha^{34}_{\D}\mathcal{W}_{\D,0}(\D_i)+\sum_{p}\dfrac{a^{12}_p\alpha^{34}_p}{m_p^2-m^2_\D}\mathcal{W}_{\D_p,0}(\D_i)+\sum_{q}\dfrac{\alpha^{12}_qa^{34}_q}{m_q^2-m^2_\D}\mathcal{W}_{\D_q,0}(\D_i). 
\ee
Here label $p,q$ denote the double-trace operator with scaling dimensions $\D_p=\D_1+\D_2+2p\,$ and with $\,\D_q=\D_3+\D_4+2q$. 
The reader who are interested in the detail of coefficients(\,$a^{ij}_k,\alpha^{ij}_k,\,$and $m^2_k$), see around (4.21) of \cite{Hijano:2015zsa}. 

When $D_{11}^n$ acts on this expression, each GWD of r.h.s. becomes GWD with an external spin-$n$ field from the results of previous subsection. One can notice that l.h.s. also becomes Witten diagram with an external spinning field: 
\begin{align}
(-2)^n D_{11}^n\mathcal{A}^{\textrm{Exch}}_4(\D_i;\D)&=\int_{\textrm{AdS}}\,dY_2 \left[\int_{\textrm{AdS}}dY_1\, G^{0}_{b\p}(X_2,Y_1;\D_2)\right.\nn\\
&\times G^{n}_{b\p}(X_1,Y_1;\D_1-n)^{A_1\cdots A_n}\left.\dfrac{\p}{\p Y^{A_1}_1}\cdots\dfrac{\p}{\p Y^{A_n}_1}G_{bb}(Y_1,Y_2;\D)\right]\nn\\
&\times G^{0}_{b\p}(X_3,Y_2;\D_3)G^{0}_{b\p}(X_4,Y_2;\D_4)\nn\\
&\equiv\mathcal{A}^{(n,0,0,0)}_4(\widetilde{\D}_i;\D),
\end{align}
where $\widetilde{\D}_i\equiv\D_i-n\delta_{i,1}$. The key identity to show above equation is 
\begin{align}
&\hspace{-5mm}D_{11}^{n}[G^{0}_{b\p}(X_1,Y_1;\D_1)G^{0}_{b\p}(X_2,Y_1;\D_2)]\nn\\
&=\left(\dfrac{1}{2}\right)^{n}\dfrac{\p}{\p Y^{A_1}_1}\cdots\dfrac{\p}{\p Y^{A_n}_1}\left[G^{n}_{b\p}(X_1,Y_1;\widetilde{\D}_1)_{A_1\cdots A_n}G^{0}_{b\p}(X_2,Y_1;\D_2)\right]. 
\end{align}
We should stress that bulk coordinate $Y_1$ is not restricted to geodesics. 
To summary, we have shown that scalar-exchange Witten diagram with an external spin-$n$ field can be written as the sum of GWD with an external spin-$n$ field. 
\begin{align}
\mathcal{A}^{(n,0,0,0)}_4(\widetilde{\D}_i;\D)&=\alpha^{12}_{\D}\alpha^{34}_{\D}\mathcal{W}^{(n,0,0,0)}_{\D,0}(\widetilde{\D}_i)\nn\\
&+\sum_{p}\dfrac{a^{12}_p\alpha^{34}_p}{m_p^2-m^2_\D}\mathcal{W}^{(n,0,0,0)}_{\D_q,0}(\widetilde{\D}_i)+\sum_{q}\dfrac{\alpha^{12}_qa^{34}_q}{m_q^2-m^2_\D}\mathcal{W}^{(n,0,0,0)}_{\D_q,0}(\widetilde{\D}_i).
\end{align}
)\fi
\section{Summary and discussion}\label{sec:summary}
In this paper, we have explicitly constructed the amplitude of the scalar exchange geodesic Witten diagrams \eqref{eq:efcpwn} that have an external field with spin and three external scalar fields. We have also found the three point interaction \eqref{eq:nint} in the bulk to construct our amplitude of GWD.
Moreover, up to normalization, we have shown that these GWD are equivalent to the conformal partial waves that also have an external field with spin. 
There are two ways to prove this equivalence: comparing GWD with the known expression of CPW and checking that GWD satisfies the conformal Casimir equation. 
In the spin-1 case, we have proven the correspondence in both ways. 
In the spin-$n$ case, we have expressed our construction of GWD as scalar GWD with differential operators. We have confirmed that this expression is just the same one as CPW in \cite{Costa:2011dw}. 

We expect that our results can be generalized to the case of any symmetric-traceless fields. One can rewrite formula (3.40) of \cite{Costa:2011dw} in terms of the scalar GWD as
\be
W^{(\ell_1, \ell_2, \ell_3, \ell_4)}_{\D,\ell}= \mathcal{D}_{12}\mathcal{D}_{34}\mathcal{W}^{(0,0,0,0)}_{\D,\ell}, 
\ee
where $\mathcal{D}_{12}$ denotes the combination of $D_{11}, D_{12}, D_{21}, D_{22}$ and $H_{12}$ introduced in \cite{Costa:2011dw}, and so dose $\mathcal{D}_{34}$. 
In this paper, we have concentrated only on CPW whose tensor structure is unique. However, it is not true in the general cases. Since the three point functions in CFT have degrees of freedom of the tensor structure of more than one in general, the tensor structure of CPW can be different even if these have the same $(\D_i,\ell_i)$ and $(\D,\ell)$. 
This should also be true in the bulk picture; hence, we need an explicit dictionary between the tensor structures of CPW and three point interactions in the bulk. 

We have studied only GWD with an external symmetric traceless tensor. 
However, CPW can contain a mixed-symmetry tensor structure in general. One can use index-free notation to construct such CPW by using Grassmann auxiliary fields introduced in \cite{Costa:2014rya, Costa:2016hju}. 
In our paper, we have rewritten $\mathcal{W}^{(n,0,0,0)}_{\D,0}$ in terms of the scalar GWD $\mathcal{W}^{(0,0,0,0)}_{\D,0}$ with differential operators acting on it. Thus $\mathcal{W}^{(0,0,0,0)}_{\D,0}$ plays the role of a seed of GWD. 
The notion of a seed was introduced in the context of CPW in \cite{Echeverri:2015rwa}. 
If one considers CPW with mixed-symmetry tensors, generally, the seed CPW include other CPW than the scalar CPW.
Therefore it is important to find explicit forms of the seed GWD that correspond to such seed CPW. 

One of the motivations to consider the bulk representation of CPW is for the new expression of the stress tensor CPW . This expression gives us the universal information about CFT and quantum gravity on AdS. For this purpose, we need to complete the above construction of GWD and we leave it for future work. 

Recently, the authors of \cite{Czech:2016xec} have proposed a bulk dual of the OPE block (see also \cite{deBoer:2016pqk}) . This bulk dual is an operator smeared over the subspace and invariant under isometry. If the points in the OPE block are spacelike, the bulk dual operator is smeared over the geodesics between these points. Since CPW can be written as the ``two point function'' of the OPE blocks, this proposal can explain why GWD corresponds to CPW. 
Their arguments are mainly based on the Lorentzian CFT; therefore, one can derive the Witten diagram representation of CPW even in the Lorentzian CFT.  
It is interesting to consider this correspondence with the spinning field. 

\bigskip
\goodbreak
\centerline{\bf Acknowledgments}
\noindent
We would like to thank N.~Iizuka, H.~Mori and S.~Yamaguchi for useful comments and discussions. The work of M.N.~was supported in part by the Grant-in-Aid for JSPS Fellows Grant Number JP15J01666.

\appendix
\section{Useful formulas for the calculations}\label{app:formula}
In this appendix, we note useful formulas that make the calculations in the paper easier.

\subsection{Formulas for section \ref{sec:1000}} 
These formulas are convenient for the proof of (\ref{eogwd}):
\begin{align}
u(\lambda)^2+|x(\lambda)-x_1|^2&=\frac{|x_{12}|^2e^{\lambda}}{2\cosh{\lambda}},\\
u(\lambda)^2+|x(\lambda)-x_2|^2&=\frac{|x_{12}|^2e^{-\lambda}}{2\cosh{\lambda}},\\
\frac{\p x^b(\lambda)}{\p x^a_1}&=\frac{\delta^b_ae^{-\lambda}}{2\cosh{\lambda}},\\
\frac{\p x^a(\lambda)}{\p \lambda}&=-\frac{(x_{12})^a}{2\cosh^2{\lambda}},\\
\frac{\p u(\lambda)}{\p x^a_1}&=-\frac{(x_{12})_a}{|x_{12}|^2}\frac{\p u(\lambda)}{\p \lambda}+\frac{(x_{12})_ae^{-\lambda}}{|x_{12}|\cosh^2{\lambda}}.
\end{align}

\subsection{Formulas for section \ref{sec:spinn}} 
In our convention, the coordinate $Y_{1}(\la)$ of the geodesic between the boundary points $X_1$ and $X_2$ is written as
\be
Y_{1A}(\la)=\dfrac{e^{-\la}X_{1A}+e^{\la}X_{2A}}{\sqrt{-2X_1\cdot X_2}}.
\ee
With $Z_i\cdot X_i=X_i\cdot X_i=Z_i\cdot Z_i=0\;(i=1,2)$, one can easily show
\begin{align}
(-2X_1\cdot X_2)&=(-2X_1\cdot Y_1(\la))(-2X_2\cdot Y_1(\la)), \\
(Z_1\cdot Y_1(\la))(X_1\cdot X_2)&=(X_1\cdot Y_1(\la))(Z_1\cdot X_2), \\
(2Z_1\cdot X_2)&=(2Z_1\cdot Y_1(\la))(-2Y_1(\la)\cdot X_2), \\
(2X_1\cdot Z_2)&=(2Z_2\cdot Y_1(\la))(-2Y_1(\la)\cdot X_1).
\end{align}
For integrating by parts, it is also useful to note that
\be
Z_1\cdot Y_1(\la)=Z_1\cdot\dfrac{dY_1}{d\la}, \;Z_2\cdot Y_1(\la)=-Z_2\cdot\dfrac{dY_1}{d\la}.
\ee
As for the scalar function $f(Y_1(\la))$, one can show
\begin{align}
X_1\cdot\dfrac{\p f(Y_1(\la))}{\p X_1}&=\dfrac{1}{(-2X_1\cdot Y_1(\la))}\,(X_1\cdot\nabla_{Y_1})f(Y_1(\la)), \\
Z_1\cdot\dfrac{\p f(Y_1(\la))}{\p X_1}&=\dfrac{1}{(-2X_1\cdot Y_1(\la))}\,(Z_1\cdot\nabla_{Y_1})f(Y_1(\la)). 
\end{align}
One can also show similar formulas with $X_1\rightarrow X_2$ and $Z_1\rightarrow Z_2$.
To check these formulas, we use $Y_{1A}\frac{\p Y^A_1}{\p X^B}=0$, $Y_1(\la)\cdot\nabla_{Y_1}=0$, and
\begin{align}
\left(\dfrac{\p Y^A_1(\la)}{\p X^B_1}\right)&=\dfrac{1}{(-2X_1\cdot Y_1(\la))}\left(\delta^A_B+\dfrac{Y^A_1(\la)X_{2B}}{(-2X_2\cdot Y_1(\la))}\right), \\
\left(\dfrac{\p Y^A_1(\la)}{\p X^B_2}\right)&=\dfrac{1}{(-2X_2\cdot Y_1(\la))}\left(\delta^A_B+\dfrac{Y^A_1(\la)X_{1B}}{(-2X_1\cdot Y_1(\la))}\right). 
\end{align}
Formulas for the derivative with respect to $\la$ are
\begin{align}
\dfrac{d}{d\la}G_{bb}(Y_1(\la), Y_2; \D)&=\dfrac{1}{X_1\cdot Y_1(\la)}(X_1\cdot\nabla_{Y_1})\,G_{bb}(Y_1(\la), Y_2; \D), \\
\dfrac{d}{d\la}\left(G^0_{b\p}(Y_1(\la),X_1; \D_1)\,G^0_{b\p}(Y_1(\la),X_2; \D_2)\right)&=(\D_2-\D_1)\,G^0_{b\p}(Y_1(\la),X_1; \D_1)\,G^0_{b\p}(Y_1(\la),X_2; \D_2).
\end{align}
To see the equivalence of GWD between section \ref{sec:1000} and section \ref{sec:spinn}, it is useful to use the induced AdS metric
\be
G_{AB}(Y)\equiv\eta_{AB}+Y_AY_B
\ee
and the relation
\be
\eta^{AB}\nabla_{B}=G^{AB}\nabla_{B}. 
\ee
\section{The three point scalar geodesic Witten diagram}\label{app:3pt}
\begin{figure}[tbp]
\begin{center}
\resizebox{60mm}{!}{\includegraphics{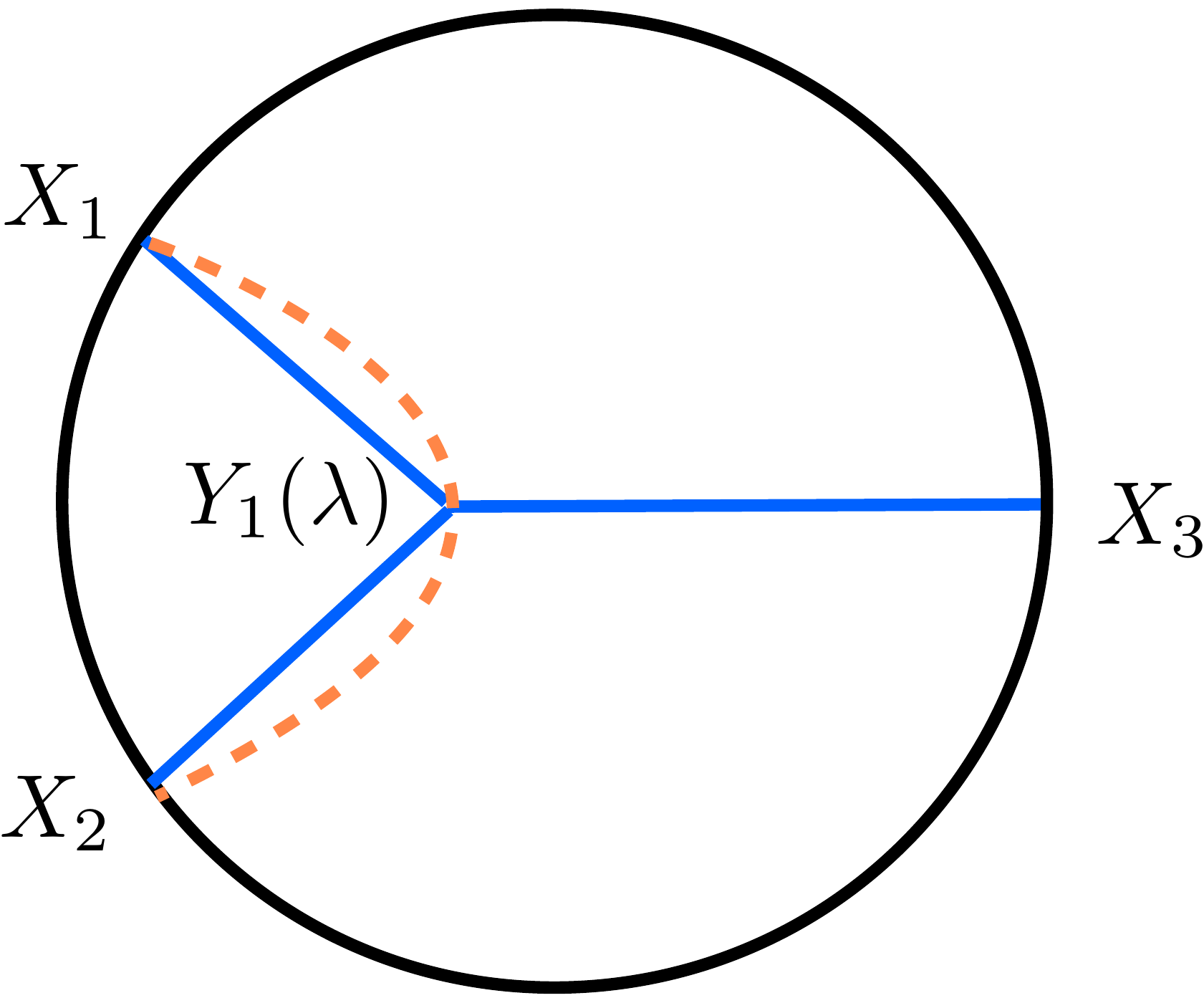}}
\caption{The three point scalar geodesic Witten diagram. The amplitude of this diagram also becomes the form of the scalar three point function in CFT as  well as the Witten diagram. }\label{tgwd}
\end{center}
\end{figure}

It is well known that the conformal three point function can be obtained from the three point Witten diagram integrated over all points in the bulk. In this appendix, we note that the three point scalar geodesic Witten diagram (Figure \ref{tgwd}) also corresponds to the conformal three point function.

If we choose the geodesic between $X_1$ and $X_2$, the amplitude of the three point scalar geodesic Witten diagram $\mathcal{W}(X_1,X_2,X_3;\D_i)$ can be defined as
\begin{align}
\mathcal{W}(X_1,X_2,X_3;\D_i)&\equiv\int^{\infty}_{-\infty} d\la\,G^0_{b\p}(Y_1(\la),X_1; \D_1)\,G^0_{b\p}(Y_1(\la),X_2; \D_2)\,G^0_{b\p}(Y_1(\la),X_3; \D_3)\nn\\
&=(-2X_1\cdot X_2)^{-\frac{1}{2}(\D_1+\D_2-\D_3)}\,(-2X_1\cdot X_3)^{-\D_3}\,\int^{\infty}_{-\infty} d\la\,\dfrac{e^{\la(\D_2-\D_1+\D_3)}}{(1+ae^{2\la})^{\D_3}}, 
\end{align}
where we define $a\equiv\frac{-2X_2\cdot X_3}{-2X_1\cdot X_3}$. 
Changing the integral variable to $u=\frac{1}{1+ae^{2\la}}$, we obtain
\begin{align}
\mathcal{W}(X_1,X_2,X_3;\D_i)
&=\dfrac{1}{2}\,B\left(\frac{1}{2}(\D_3+\D_1-\D_2),\,\frac{1}{2}(\D_2+\D_3-\D_1)\right)\nn\\
&\times\dfrac{1}{(-2X_1\cdot X_2)^{\frac{1}{2}(\D_1+\D_2-\D_3)}}\dfrac{1}{(-2X_2\cdot X_3)^{\frac{1}{2}(\D_2+\D_3-\D_1)}}\dfrac{1}{(-2X_3\cdot X_1)^{\frac{1}{2}(\D_3+\D_1-\D_2)}}.\label{eq:3ptgwd}
\end{align}
Here $B(x,y)$ is the beta function $B(x,y)\equiv\int^1_0 du u^{x-1}(1-u)^{y-1}=\frac{\G(x)\G(y)}{\G(x+y)}$ and this integral is convergent if $\Re x>0, \Re y>0$. 
For the convergence of the amplitude of GWD \eqref{eq:3ptgwd}, we need the condition
\be
\D_3>|\D_1-\D_2|, 
\ee
which is the same condition in section \ref{sec:1000}. After substituting $(-2X_i\cdot X_j)=|x_{ij}|^2$ into \eqref{eq:3ptgwd}, it is the same as \eqref{stpf} up to constant. In other words, the three point scalar GWD also provides the three point scalar correlation function in CFT. We note that (\ref{rtps}) can be rewritten in terms of the three point GWD by replacing  $G_{bb}(y(\lambda), y(\lambda'); \Delta)$ with $G_{b\partial}(y(\lambda), x_3; \Delta_3)$ in (\ref{e311}).

\section{Proof by embedding formalism}\label{sec:emb}
In this appendix, we give a more transparent proof of the correspondence discussed in section \ref{sec:1000} by using the embedding formalism and the conformal Casimir equation. The embedding formalism is reviewed in section \ref{subsec:emb}. 
\subsection{Conformal Casimir equation}
There is another way to show the correspondence between GWD $\mathcal{W}^{(1, 0, 0, 0)}_{\D, 0}(x_i; \D_i)$ and CPW $W^{(1, 0, 0, 0)}_{\D, 0}(x_i; \D_i)$, namely, checking that GWD satisfies the conformal Casimir equation. 
This is because the conformal Casimir equation is the equation of which CPW is the solution. 
First, we derive the conformal Casimir equation for $W^{(1, 0, 0, 0)}_{\D, 0}(x_i; \D_i)$. 

For convenience, we introduce the Lorentz generators $L_{AB}$ in $(d+2)$-dimension which are equivalent to the generators of conformal symmetry $SO(d+1,1)$ in $d$-dimension. Any local field $\mathcal{O}(x)$ in CFT is transformed under $L_{AB}$ as
\be
[L_{AB}, \mathcal{O}(x)]=(L_x)_{AB}\mathcal{O}(x). \label{eq:ctransf}
\ee
Here $L_x$ is the differential operator acting on fields at $x$ and $AB$ denotes the label of the generators. The explicit form of $(L_x)_{AB}$ depends on the conformal dimension $\D$ and spin $\ell$ of $\mathcal{O}(x)$, which are now suppressed. We will display $(L_x)_{AB}$ explicitly after introducing the embedding formalism. 

In terms of the complete set for the conformal family, CPW $W^{(1,0,0,0)}_{\D,\ell}$ can be expressed as
\be
W^{(1,0,0,0)}_{\D,0}=\dfrac{1}{C_{12\mathcal{O}}C_{34}{}^{\mathcal{O}}}\sum_{\alpha}\bra{0}\mathcal{J}_1(x_1)\mathcal{O}_2(x_2)\ket{\alpha}\bra{\alpha}\mathcal{O}_3(x_3)\mathcal{O}_4(x_4)\ket{0}, 
\ee
where $\ket{\alpha}$s denote the scalar primary state $\ket{\mathcal{O}}$ whose conformal dimension is $\D$ and its descendants. In order to derive the conformal Casimir equation for CPW, we define the quadratic Casimir $L^2\equiv \frac{1}{2}L_{AB}L^{AB}$. 
The scalar primary state $\ket{\mathcal{O}}$ has  the eigenvalue $C_2(\D,0)$ of $L^2$\cite{Dolan:2000ut}, 
\be
L^2\ket{\mathcal{O}}=C_2(\D,0)\ket{\mathcal{O}}=-\D(\D-d)\ket{\mathcal{O}}. 
\ee
Since $L^2$ commutes with all generators $L_{AB}$, $L^2$ has the same eigenvalue in the descendants. 
 
Let us consider how $L^2$ acts on CPW. From (\ref{eq:ctransf}) and the conformal invariance of the vacuum, we obtain
\begin{align}
(L^{(1)}_{x_1}+L^{(0)}_{x_2})_{AB}\bra{0}\mathcal{J}_1(x_1)\mathcal{O}_2(x_2)\ket{\alpha}=-\bra{0}\mathcal{J}_1(x_1)\mathcal{O}_2(x_2)L_{AB}\ket{\alpha}, \label{eq:wt3pt}
\end{align}
where $\ell$ of $L^{(\ell)}_{x}$ implies spin of the operator on $x$. 
Using (\ref{eq:wt3pt}) twice, we get 
\begin{align}
(L^{(1)}_{x_1}+L^{(0)}_{x_2})^2\bra{0}\mathcal{J}_1(x_1)\mathcal{O}_2(x_2)\ket{\alpha}&=\bra{0}\mathcal{J}_1(x_1)\mathcal{O}_2(x_2)L^2\ket{\alpha}.
\end{align}
Since all of $\ket{\alpha}$s have the same Casimir eigenvalue $C_2(\D,0)$, we obtain the second-order differential equation
\be
(L^{(1)}_{x_1}+L^{(0)}_{x_2})^2 W^{(1,0,0,0)}_{\D,0}=C_2(\D,0)W^{(1,0,0,0)}_{\D,0}. \label{eq:casimir}
\ee
This equation is the so-called conformal Casimir equation for CPW. After taking the appropriate boundary condition, we will obtain a unique solution (up to constant). One can easily extend the above discussion to the case of generic external primary operators and intermediate states. 
\subsection{Proof by the conformal Casimir equation}
Based on the above preparation, we show that GWD $\mathcal{W}^{(1,0,0,0)}_{\D,0}$ satisfies the conformal Casimir equation \eqref{eq:casimir}. 
Any field $\mathcal{O}$ that belongs to any spin representation on the projective light cone is transformed by the generators $L_{AB}$ as
\be
[L_{AB}, \mathcal{O}(X;Z)]=(L^{(\ell_{\mathcal{O}})}_X)_{AB}\mathcal{O}(X;Z), \label{eq:sod+11}
\ee
where
\be
(L^{(\ell_{\mathcal{O}})}_X)_{AB}\equiv X_A\dfrac{\p}{\p X^B}-X_B\dfrac{\p}{\p X^A}+S^{(\ell_{\mathcal{O}})}_{AB}. 
\ee
Here $S^{(\ell_{\mathcal{O}})}_{AB}$ depends on spin $\ell_{\mathcal{O}}$ of $\mathcal{O}(X;Z)$, for example, 
\be
S^{(0)}_{AB}=0, \hspace{3mm} (S^{(1)}_{AB})_{CD}=\eta_{AC}\eta_{BD}-\eta_{BC}\eta_{AD}. 
\ee
We define $S^{(1)}_{AB}$ as $(S^{(1)}_{AB})_{CD}$ in the index-free notation\footnote{We use $\frac{\p}{\p Z^A}$ instead of $D_A$ introduced in \cite{Costa:2011mg} because these are effectively the same operators when acting on first-order polynomials of $Z^A$.}, 
\be
S^{(1)}_{AB}\equiv Z_C (S^{(1)}_{AB})^{CD}\dfrac{\p}{\p Z^D}=Z_A\dfrac{\p}{\p Z^B}-Z_B\dfrac{\p}{\p Z^A}.
\ee

Since the isometry group of AdS is also $SO(d+1,1)$, we can use the same operators for generators of isometry. For example, we can define a differential operator for isometry, 
\be
(L^{(0)}_Y)_{AB}\equiv Y_A\dfrac{\p}{\p Y^B}-Y_B\dfrac{\p}{\p Y^A}.
\ee
It is enough to define $L^{(0)}_Y$ because we concentrate on scalar exchange in the bulk. We will use an important identity\cite{Balasubramanian:1998sn}
\be
-\dfrac{1}{2}(L^{(0)}_Y)_{AB}(L^{(0)}_Y)^{AB}\,f(Y)=\nabla^2_Yf(Y), \label{eq:Lnabla} 
\ee
where $f(Y)$ is an arbitary scalar function on AdS.

Let us check that GWD $\mathcal{W}^{(1,0,0,0)}_{\D,0}$ satisfies the conformal Casimir equation. 
Our $F^{(1,0,0)}_{\widetilde{\D}_1,\widetilde{\D}_2,\D}(X_1,X_2,Y_2;Z_1)$ is manifestly invariant under $SO(d+1,1)$ rotation. This means
\be
(L^{(1)}_{X_1}+L^{(0)}_{X_2}+L^{(0)}_{Y_2})_{AB}\,F^{(1,0; 0)}_{\widetilde{\D}_1,\widetilde{\D}_2,\D}(X_1,X_2,Y_2;Z_1)=0. \label{eq:wtF}
\ee
By using \eqref{eq:Lnabla} and \eqref{eq:wtF}, we obtain a key identity
\be
-(L^{(1)}_{X_1}+L^{(0)}_{X_2})^2\,F^{(1,0; 0)}_{\widetilde{\D}_1,\widetilde{\D}_2,\D}=\nabla^2_{Y_2}\,F^{(1,0; 0)}_{\widetilde{\D}_1,\widetilde{\D}_2,\D}.
\ee
Since $G_{bb}(Y_1, Y_2; \D)$ is an eigenfunction of $\nabla^2_{Y_2}$ and its eigenvalue is $\D(\D-d)$\cite{Fronsdal:1974ew}, we obtain
\be
-(L^{(1)}_{X_1}+L^{(0)}_{X_2})^2\,\mathcal{W}^{(1,0,0,0)}_{\D,0}=\D(\D-d)\mathcal{W}^{(1,0,0,0)}_{\D,0},
\ee
where we have assumed that the two geodesics do not intersect each other \cite{Hijano:2015zsa}. 
This is just the same as the conformal Casimir equation for CPW. Thus we have shown that GWD $\mathcal{W}^{(1,0,0,0)}_{\D,0}$ satisfies the conformal Casimir equation and GWD $\mathcal{W}^{(1,0,0,0)}_{\D,0}$ is equivalent to CPW $W^{(1,0,0,0)}_{\D,0}$ in the embedding formalism.

\end{document}